\documentclass{article}
\newcommand{\A}{\mathcal{A}}
\newcommand{\lev}{{*\!R}_{\alpha\beta\mu\nu}}
\newcommand{\pr}{{R\!*}_{\alpha\beta\mu\nu}}
\newcommand{\dv}{{*\!R\!*}_{\alpha\beta}{}^{\mu\nu}}
\newcommand{\er}{R_{\alpha\beta\mu\nu}}
\newcommand{\es}{\mathcal{S}_{\alpha\beta\mu\nu}}
\newcommand{\err}{\mathcal{R}_{\alpha\beta\mu\nu}}
\newcommand{\dva}{{*\!R\!*}_{\alpha\beta\mu\nu}}
\begin{document}
\title{On General Relativity extension\thanks{University of Petrozavodsk, Russia}}%
\author{A. L. Koshkarov\thanks{email: Koshkarov@petrsu.ru}}%
\date{}%
\begin{abstract}
The thorough analysis of the duality properties of the Riemann curvature
tensor points  to possibility of extension of Einstein's General Relativity
to the nonabelian Yang-Mills theory. The motion equations of the theory
are Yang-Mills' equations for the curvature tensor. Einstein's equations
(with cosmological term to appear as an integration constant) are
contained in the theory proposed. New is that now gravitational field
is {\em not} exceptionally determined by  matter energy-momentum but can
possess its own {\em non-Einsteinian} dynamics (vacuum fluctuations,
self-interaction) which is
generally an attribute  of nonabelian gauge field.
The gravitational equations proper due to
either matter energy-momentum or vacuum fluctuations are side conditions
imposed on the Riemann tensor, like self-duality conditions. One of such
conditions in the end
results in Einstein's equations, other ones are the gravitational instantons
equations.

\end{abstract}
\maketitle
\section{Introduction}

There is no doubt that Einstein's General Relativity  is nonabelian
gauge theory although not quite  similar to conventional the Yang-Mills
theory.

Nevertheless, there are rather too much arguments in favor of the theory is
nonabelian. But in what way, the fact that gravitation is nonabelian does
get on with widely spread and prevailing view the gravity source is
 energy-momentum and only energy-momentum?
 And how about nonabelian self-interaction? Of course, here we touch very
 tender spots about exclusiveness of gravity as physical field, the energy
 problem, etc. Still the spherically-symmetric field out of Schwarzschild's
 sphere  looks quite like
 Coulomb's solution in Electrodynamics, the abelian theory without
 self-interaction.
 All the facts point out the General Relativity is not quite conventional
 nonabelian theory. In addition, Einstein's equations are not like
  Yang-Mills'.

It is shown in this paper the theory can be formulated {\sl ad exemplum}
ordinary Yang-Mills' theory with more or less standard description
in the  form of the Yang-Mills equation, with self-interactions and instantons.
For all that, Einstein's equations are contained in the theory rather than
cancelled and do not dwindle. And their existence as themselves seems to
relate to the gravity peculiarities.

For our purposes,
it is  essential  the fact that internal (group) indices and
space-time one are interchangeable, i.~e. group acts in the Minkowski
space-time which as a result get curved. In fact the internal space  coincides
with the space-time. Therefore, it is convenient  to hold the viewpoint  that
the first  two indices  $\alpha,\beta$ of the curvature tensor $\er$ are
internal, and the second pair $\mu,\nu$  are the spacetime indices.
And vice versa that's right as well. This is the peculiar features of
the gravity as the gauge theory. For this reason the gravity duality
properties
are even more nontrivial and interesting than those in the ordinary
Yang-Mills theory.

\section{The duality properties of the Riemann tensor}

The duality properties us to be interested in have been established in the
article~\cite{dua} which though includes some mistakes.
About notations.
The metric  with signature $(+,-,-,-)$ in $D=4$
pseudo-riemannian manifold is given to be
metric-compatible to (Riemannian) connection
in the regular way.

Let us introduce operations: 1) the left dual conjugation
($\lev$),
2) the right dual conjugation
($\pr$),
and  3) twice dual conjugation
($\dv$)
$$
\lev=\frac{1}{2}E_{\alpha\beta\rho\sigma}R^{\rho
\sigma}{}_{\mu\nu},\
\pr=\frac{1}{2}R_{\alpha\beta}{}{}^{\rho\sigma}
E_{\rho\sigma\mu\nu},\
\dv=\frac{1}{2}E_{\alpha\beta\rho\sigma}
R^{\rho\sigma}{}_{\gamma\delta}E^{\gamma\delta\mu\nu},
$$
where
$
 E_{\alpha\beta\mu\nu}=\sqrt{-
g}\varepsilon_{\alpha\beta\mu\nu}
$ -
the Levi-Civita tensor,
 $g$ - the metric tensor determinant.

For example
$$
*\!{*\!R}_{\alpha\beta\mu\nu}={{R\!*}\!*}_{\alpha\beta\mu\nu}=-R_{\alpha\beta\mu\nu}
$$
It is usual properties of double  dual conjugates in the
(+,-,-,-) riemannian space.

In terms of the dual conjugates the cyclicity identity
$$
R_{\alpha\beta\mu\nu}+R_{\alpha\nu\beta\mu}+R_{\alpha\mu\nu\beta}=0
$$
is of the form
$$
{*\!R}_\alpha{}_{\mu}\equiv{*\!R}_\alpha{}^\nu{}_{\mu\nu}=0\mbox{ and/or }
{R\!*}_\alpha{}{}_{\mu}\equiv{R\!*}_\alpha{}^\nu{}_{\mu\nu} =0,
$$
Bianchi's identity
$$
R_{\mu \nu \rho \sigma ;\delta }+R_{\mu \nu \delta \rho ; \sigma }+
R_{\mu \nu \sigma \delta ; \rho }=0
$$
transforms to
$$
*\!R_{\mu}{}^{\nu}{}_{\rho\sigma ; \nu }=0 \mbox{  and/or }
{R\!*}_{\mu \nu \rho }{}{}{}^\sigma {}_{;\sigma }=0
$$
Or that can be rewritten as follows
$$
{*\!R\!*}_{\alpha\beta\mu}{}^\nu{}_{;\nu} =0\mbox{ and/or}\,
{*\!R\!*}_\alpha{}^\beta{}_{\mu\nu;\beta}  =0
$$
And twice dual conjugate Riemann's tensor
 $\dv$
 can be represented by Riemann's and its contractions, i.~e. by Ricci's tensor
 and scalar, for the expression
$$
E^{\alpha\beta\rho\sigma}E_{\gamma\delta\mu\nu}=
\varepsilon^{\alpha\beta\rho\sigma}\varepsilon_{\gamma\delta\mu\nu}
$$
can be calculated and expressed by the Kronecker $\delta$-symbols~\cite{dua}:
\begin{equation}\label{1}
\dv=-R^{\mu\nu}{}_{\alpha\beta}+
\delta^\mu_\alpha R^\nu_\beta+\delta^\nu_\beta R^\mu_\alpha
-\delta^\mu_\beta R^\nu_\alpha-\delta^\nu_\alpha R^\mu_\beta
-\frac{1}{2}R\delta^{\mu\nu}_{\alpha\beta}=
\end{equation}
Hereafter the notations are used
$$
\delta^{\mu\nu}_{\alpha\beta}=\delta_\alpha^\mu\delta_\beta^\nu-
\delta_\alpha^\nu\delta_\beta^\mu ,\quad
g_{\alpha\beta\mu\nu}=g_{\alpha\mu}g_{\beta\nu}-
g_{\alpha\nu}g_{\beta\mu}
$$
The next important step is to expand the Riemann tensor into sum of two parts,
\begin{equation}\label{2}
\er=\frac{1}{2}(\er+\dva+\er-\dva)=\err+\es
\end{equation}
where
\begin{equation}\label{3}
\err=\frac{1}{2}(\er-\dva),\quad\es=\frac{1}{2}(\er+\dva)
\end{equation}
Now one can represent the tensors $\err$  and $\es$
by Riemann's tensor and Ricci's tensor and scalar
\begin{equation}\label{4}
\err=\er-\frac{1}{2}(g_{\alpha\mu}R_{\beta\nu}+g_{\beta\nu}R_{\alpha\mu}
-g_{\alpha\nu}R_{\beta\mu}-g_{\beta\mu}R_{\alpha\nu}-
\frac{1}{2}Rg_{\alpha\beta\mu\nu})
\end{equation}
\begin{equation}\label{5}
\es=\frac{1}{2}(g_{\alpha\mu}R_{\beta\nu}+g_{\beta\nu}R_{\alpha\mu}
-g_{\alpha\nu}R_{\beta\mu}-g_{\beta\mu}R_{\alpha\nu}-
\frac{1}{2}Rg_{\alpha\beta\mu\nu})
\end{equation}

One must already say  something about the tensors properties. The
$\es$ is  noteworthy. Note it is expressed by the Ricci tensor and scalar
only, not by Riemann's.

The tensor $\err$ should not be confused with the Weyl conformal  tensor
 $C_{\alpha\beta\mu\nu}$
$$
\err=C_{\alpha\beta\mu\nu}+\frac{1}{12}Rg_{\alpha\beta\mu\nu}
$$

Further, when twice dual conjugating, both $\es$ and $\err$
transform  {\em simply}
\begin{equation}\label{6}
{*\!\mathcal{S}\!*}_{\alpha\beta\mu\nu}=+\es,\quad
{*\!\mathcal{R}\!*}_{\alpha\beta\mu\nu}=-\err
\end{equation}
i.~e., $\es$ and $\err$   are respectively twice selfdual and antiselfdual
parts of the curvature tensor.

It makes sense  to introduce  a new "quantum" number --- d-parity,
characterizing behavior of tensors (like curvature one) under twice dual
conjugation. For example, $\err$ is odd , and $\es$ -
even under d-parity reflection.
 Two more examples of d-odd tensors are $g_{\alpha\beta\mu\nu}$
and $E_{\alpha\beta\mu\nu}$.

There are nontrivial equations
\begin{equation}\label{7}
\es=0
\end{equation}
and
\begin{equation}\label{8}
\err=0
\end{equation}
These equations have a direct relationship to instantons in nonabelian
gauge theories. In particular in the case of $SO(4)$ or  $SU(2)$ gauge
group, they describe Belavin-Polyakov-Schwarz-Tyupkin instanton and
antiinstanton~\cite{bpst}.

Below we shall see the equation~(\ref{8}) describes the gravitational
 instantons.

Some solutions to these equations have been obtained in ~\cite{dua} .
For example, the equation~(\ref{7}) has a static solution in the metric
\begin{equation}\label{9}
ds^2=e^{\nu(t,r)}dt^2-e^{\lambda(t,r)}dr^2-r^2(d\theta^2+\sin^2\theta d\phi^2)
\end{equation}
Six equations~(\ref{7}) with nonvanishing left member reduce to the only
second order equation
$$
\lambda =-\nu ,\qquad \nu ''(r)+\nu '^2(r)=\frac{2}{r^2}(1-e^{-\nu })
$$
The solution is
\begin{equation}\label{10}
e^\nu =1+C_1r^2+\frac{C_2}{r}
\end{equation}

Thus central-symmetric solution to equation~(\ref{7}) is static and quite
similar to Schwarzschild's except for $C_1r^2$.
It is not without purpose and  we'll be back
to this as well as to equations~(\ref{7}) and~(\ref{8}). Below we'll see
that for  equation~(\ref{7}) with vanishing right hand side
$C_2=0$.

\section{From  Einstein's  to gravitational Yang-Mills' equations }

Solution to equation~(\ref{7}) including the Schwarzschild solution
suggests that it is  possible to use this equation instead of Einstein's
\begin{equation}\label{11}
R_{\alpha\mu}- \frac{1}{2}Rg_{\alpha\mu}=\Lambda g_{\alpha\mu}+T_{\alpha\mu},
\quad R+T=-4\Lambda
\end{equation}
although in emptiness. Really, solution to this equation in the metric~(\ref{9})
 coincide with~(\ref{10}),  if $C_1=\Lambda$. Even more so, the tensor $\es$
 in the left hand side~(\ref{7})  is fully determined by the Ricci tensor.

At this point, we want to call attention to one  little drawback to the Einstein
equations. Of course, at times, there had been treating
various advantages and disadvantages of these equations although the former
are of the overwhelming majority. The one had most likely been discussed
before.
The point is that~(\ref{11}) are  system of differential equations  of the
second order in metric. And the Scwarzschild solution has merely one integration
constant. When more attentive treating Schwarzschild's problem, it turns out
 that among the equations there are both first order equations and
second order those. With all that, solutions to first order equations
and second order equations are compatible  provided that one of the two
integration constants is strictly fixed.
It is the reason that the Einstein second order equation system solution
(the Scwarzschild solution) contains solely  one integration constant.

This fact is of course known but completely ignored.
We find on this point, Einstein's equations  are somewhat inconsistent.

Further heuristically, there will be obtained equation
to generalize Einstein's equation~(\ref{11}).
Then the new equation  will be proclaimed as  one of the basic equations
describing gravity produced by matter.
After that, it will be seen the new equation implies the gravitational
Yang-Mills equation. At last, Einstein's equations will be shown to follow
from both the new equation and gravitational Yang-Mills'.

First expressing the Ricci tensor from~(\ref{11}) and substituting that
in~(\ref{5}),
 we can find
\begin{equation}\label{st}
\es=\Theta_{\alpha\beta\mu\nu}
\end{equation}
where
\begin{equation}\label{thet}
\Theta_{\alpha\beta\mu\nu}=
\frac{1}{2}(g_{\alpha\mu}T_{\beta\nu}+g_{\beta\nu}T_{\alpha\mu}
-g_{\alpha\nu}T_{\beta\mu}-g_{\beta\mu}T_{\alpha\nu}-
\frac{1}{2}Tg_{\alpha\beta\mu\nu})
\end{equation}
This  tensor is built from the metric tensor and the energy-momentum
tensor. Further it will be used "instead" of the energy-momentum tensor.
Note that the tensor $\Theta_{\alpha\beta\mu\nu}$ is d-even
like $\es$.

Now the important step follows. Let us forget the Einstein equation and instead
consider the equation~(\ref{st}) as one of the basic equations of gravity
generated by matter.

Differentiate  covariantly with respect to $x^\nu$ the equation~(\ref{st})
\begin{equation}\label{thet,n}
\mathcal{S}_{\alpha\beta\mu}{}^\nu{}_{;\nu}=
\Theta_{\alpha\beta\mu}{}^\nu{}_{;\nu}
\end{equation}
Having remembered what is $\es$~(\ref{5}), it follows that
\begin{equation}\label{r,n}
R_{\alpha\beta\mu}{}^\nu{}_{;\nu}=2\Theta_{\alpha\beta\mu}{}^\nu{}_{;\nu}
\equiv J_{\alpha\beta\mu}
\end{equation}

So, the tensor $\Theta_{\alpha\beta\mu\nu}$ determines the matter current
or the gravity matter source in the gravitational Yang-Mills equation.
 The current is not even conserved covariantly since the multiple covariant
 derivatives don't commute.

Now making contraction over $\alpha,\mu$
$$
R_{\beta}{}^\nu{}_{;\nu}=2\Theta_{\beta}{}^\nu{}_{;\nu}
$$
and taking into account the Bianchi identity and explicit form of the tensor
$\Theta_{\alpha\beta\mu\nu}$
we obtain
\begin{equation}\label{soh}
(R+T)^{,\beta}  =0
\end{equation}
 After integration
\begin{equation}\label{sohr}
R+T=-4\Lambda
\end{equation}
where $-4\Lambda$ is the integration constant.

Now we have arrived at the cross-roads. There are two alternatives.
One can consider the equation~(\ref{st}) as a basic one. Then it implies both
equations
(\ref{r,n}) and (\ref{sohr}).

Otherwise, we can consider the gravitational Yang-Mills equation~(\ref{r,n})
as a basic one. Then we have to take (\ref{st}) as a condition.
This second option is preferable.

Once again let's be back to equation~(\ref{r,n}).
 It  is of the form of the
Yang-Mills equation. The proposal is to consider it as a basic gravitational
equation. And the equality~(\ref{sohr}) is the  integral of motion, i.~e.
the conservation law.

Let's demonstrate that the Einstein equations are implied by basic
 equations~(\ref{r,n})
 and~(\ref{st}). First the conservation law~(\ref{sohr}) is obtained
from~(\ref{r,n}). Then contract~(\ref{st}) over $(\beta,\nu)$ and eliminate
$T$ by means of~(\ref{sohr}). As a result, we have exactly Einstein's
equations~(\ref{11}).
  $\Lambda$, the constant is obviously interpreted  as a
{\em cosmological term} appeared as an integration one!

Thus it is shown that the equations~(\ref{st}), (\ref{r,n}) are equivalent
to Einstein's. It is the equations those are  the basic gravitational
equations. The equation~(\ref{r,n}) is the {\em basic dynamic} one, and
another one~(\ref{st})
  is a {\em side condition} which  among fields singles out those
generated by matter.

Coming back to equations~(\ref{st}), or to~(\ref{7}) in emptiness,
we can see  the equations solution
in emptiness (\ref{10}) includes two integration constants, one of which apparently
associates with $\Lambda$.
The solution describes (out of the matter distribution) the empty constant
 curvature space with the scale factor
$1/\sqrt{\Lambda}$ and with the central-symmetry distributed about point
of origin matter.
Let a point mass be at origin. Then out of origin, the metric is given
by~(\ref{10}),
 and $C_2$ is proportional to the mass.  As for another constant,
it seems to be possible only $C_1$ proportional to $\Lambda$. Thus~(\ref{10})
is the Schwarzschild static solution in the constant curvature space.
We consider it  as a manifestation of fact that
 gravity is nonabelian.
The solution~(\ref{10}) describes  the local geometry in the neighborhood
of some spherically symmetric matter distribution.
This geometry is determined by both the mass(more precisely, energy-moment)
  and $\Lambda$.
Is that $\Lambda$ the same in case of any mass or not? In other words, is
 cosmological constant $\Lambda$
 {\em universal}?

Classically, the solution~(\ref{10}) if you wish could be interpreted as exhibition
of asymptotic freedom in gravity.

It specially should be noted that the vacuum (in emptiness) solution to
the equation~(\ref{st}), i.~e. (\ref{10}) is nontrivial, as distinct from
the Einstein theory.
That is, this solution does not just reduce to the Minkowski spacetime.
There are both static solutions~(\ref{10}) and nonstatic ones with the de Sitter
asymptotic. For example, for (closed) Robertson-Walker metric
\begin{equation}\label{rw}
ds^2=dt^2-a^2(t)\left(d\chi^2+\sin^2\chi(d\theta^2+\sin^2\theta d\phi^2)\right)
\end{equation}
the equation~(\ref{st}) in emptiness for the scale factor $a(t)$ is of the form
$$
a\ddot{a}-\dot{a}^2-1=0.
$$
The vacuum solution is
$$
a(t)=a_0\cosh\frac{t-t_0}{a_0}.
$$
Similarly, for the open metric
\begin{equation}\label{rw'}
ds^2=dt^2-a^2(t)\left(d\chi^2+\sinh^2\chi(d\theta^2+\sin^2\theta d\phi^2)\right)
\end{equation}
the equation for $a(t)$
$$
a\ddot{a}-\dot{a}^2+1=0,
$$
has a solution
$$
a(t)=a_0\sinh\frac{t-t_0}{a_0}
$$
or
$$
a(t)=a_0\sin\frac{t-t_0}{a_0}
$$
For the latter case $a(t)$ is alternating in sign that seems not to be of
physical meaning.

One can use the model equations e.~g. to construct cosmological models.
That's done. On cursory examination,  the Einstein-Friedmann cosmology
remains intact.
However now the cosmological term  seem to be the necessary element of the
theory. It should be experimentally measured in the observation cosmology.
In the sense, "the dark matter problem" might seem otherwise.
Universality of $\Lambda$-term in this approach is open to question.

\section{Non-Einsteinian gravity}

It is quite clear that the equations  (\ref{r,n}) are wider than the
Einstein General Relativity. Namely, any real gravitational fields is
considered to
obey these equations. Of all fields, the Einstein theory extracts the ones
 to be generated by the matter energy- momentum. Within the theory proposed,
 the extraction happens by imposing the side condition~(\ref{st}).
 This condition  is analogous  to self-duality conditions for instantons
 in the Yang-Mills theory. However it is not vacuum one.
It is possible to treat some other conditions which might extract
non-Einsteinian solutions for gravitational fields.

Let us try to discuss possible conditions for gravity.
Quite general side condition for equation~(\ref{r,n}) is of the form
$$
\er=\kappa\pr+\epsilon\dv+\lambda g_{\alpha\beta\mu\nu}+
\zeta E_{\alpha\beta\mu\nu}+2\Theta_{\alpha\beta\mu\nu}
$$
This is more general than~(\ref{st}). This implies the basic equation~(\ref{r,n})
 fulfilled.
In the matter presence $\Theta_{\alpha\beta\mu\nu}\neq0$, and taking
$\epsilon=-1, \kappa,\lambda,\zeta=0$, we obtain the equation~(\ref{st}).
Then the Einstein equation holds and the source conservation takes place.
Consequently, the gravitational field equations imply   motion
equations of matter in the gravitation field generated by the matter.

All that will not occur with alternative set of constants $\epsilon,
\kappa, \lambda,\zeta$ . Still admissibility of such a condition is open
to question.

In the case  of $\Theta_{\alpha\beta\mu\nu}=0$ we deal with the {\em vacuum}
side conditions.

Vacuum solutions to equation~(\ref{r,n}) could be  called gravitational instantons.
General representation for instanton is
\begin{equation}\label{gi}
\er=\kappa\pr+\epsilon{*\!R\!*}_{\alpha\beta\mu\nu}+\lambda g_{\alpha\beta\mu\nu}+
\zeta E_{\alpha\beta\mu\nu}
\end{equation}
If the condition holds then gravitational Yang-Mills equation~(\ref{thet})
will be obeyed.
The constants $\kappa,\epsilon,\lambda,\zeta$ are not quite arbitrary
and should be determined.

Alternatively instead of~(\ref{gi}), one can consider the equation
$$
\er=\kappa\lev+\epsilon\dv+\lambda g_{\alpha\beta\mu\nu}+\zeta E_{\alpha\beta\mu\nu}.
$$
Its solution provides  the equation
$$
R_\alpha{}^\beta{}_{\mu\nu;\beta}=0
$$
to be fulfilled.

The analysis of the equation~(\ref{gi}) is rather complicated so we shall
restrict our consideration to particular cases.

We have already discussed the vacuum solutions to  the equation~(\ref{st}).
Here is another possibility: $\epsilon=1;\kappa,\zeta =0$, $\lambda$ is
arbitrary.  Then we have
the non-Einsteinian equation
\begin{equation}\label{rgo}
\err=\lambda g_{\alpha\beta\mu\nu}
\end{equation}
Both left and right members of the equation are d-even.
Contracting over $\alpha,\mu$ and $\beta,\nu$, we obtain
$$
R=12\lambda
$$
Let's remind we have the matter vacuum, hence $T=0$.
It seems just in cosmology $\lambda$  to relate to the cosmology term.
Generally it is arbitrary and possibly relates to the local vacuum
fluctuations of the gravitational fields in Universe.

The equation~(\ref{rgo}) can be solved in the (closed) Robertson-Walker metric.
The equation for $a(t)$
\begin{equation}\label{a}
a\ddot{a}+\dot{a}^2+1=-2\lambda a^2
\end{equation}
has a solution
\begin{equation}\label{ra}
a(t)=\sqrt{C_1\exp(2\sqrt{-\lambda}t)+
C_2\exp(-2\sqrt{-\lambda}t)
-\frac{1}{2\lambda}}
\end{equation}
It's not analytic in $\lambda$. Note a simple constant solution
$$
a(t)=a_0=-\frac{1}{2\lambda}
$$
This solution describes the empty constant positive curvature space.
Could not it be called a gravipole?
Really it is the same solution as the static solution~(\ref{10})
 to equation~(\ref{7})
 in emptiness, i.e. with $C_2=0$. It is impossible to pass directly
on to $\lambda=0$ in this metric. However that  corresponds to  solution to
 equation~(\ref{rgo}) as the empty Minkowski space.

 In the case $\lambda=0$,
there is timedependent solution in the metric ~(\ref{rw})
$$
a(t)=a_0\sqrt{1-\left(\frac{t-t_0}{a_0}\right)^2}
$$
There are similar solutions in the open Robertson-Walker metric as well.
Interestingly, the matter motion (e.~g. the test point mass) in the vacuum
gravitational fields is already not determined  by the field equations
but obeyed the geodesic equation .

New approach  allows more directly than before to discuss
topological effects  in gravitation. Really, the conditions~(\ref{rgo})
are "topological".
Projecting~(\ref{rgo}) onto $R^{\alpha\beta\mu\nu}$ (i.e. multiplying and
contracting) results in
$$
R^{\alpha\beta\mu\nu}R_{\alpha\beta\mu\nu}-
2\lambda R=\kappa R^{\alpha\beta\mu\nu}\pr+
\epsilon R^{\alpha\beta\mu\nu}{*\!R\!*}_{\alpha\beta\mu\nu}
$$
After integration, we can see that (in case of convergence) topological
numbers can be expressed in terms of invariants quadratic and linear in
curvature. Nontrivial topological solutions seem to exist in manifolds with
Euclidean signature.

\section{Conclusion}

So, new version of gravity is proposed. It
is in form and fact the nonabelian Yang-Mills theory of gravitational field
with own rich dynamics and nontrivial topology.
The theory contains the Einstein General Relativity.

It is not impossible  to avoid a question: what are the Einstein equations?
Do they express any conservation law? Or are they  any compatibility
conditions due to the gauge group peculiarity? It should be specially noted
that in the new theory  the dynamics is described by the equation~(\ref{r,n})
 in presence of the side condition~(\ref{st}) and/or others. And the Einstein
 equations
themselves are sequel of this condition and the conservation law~(\ref{sohr}).
We have to consider various conditions   as well as Einstein's
equations as {\em constraints}. Perhaps that might change the situation with
quantizing gravity.

As for application the theory to astrophysics and cosmology, it is the next job ahead.
At once one can say that standard Einstein-Friedmann cosmological model seems not to change.
Cosmological term situation may get more definitive. It is not matter of a taste:
to work or not to work with that. One must to measure that. Once again one have to say that it is
not clear whether $\Lambda$ is universal.

Nothing to say as yet about black hole physics  in new approach.
Task in hand  is to search for the nontrivial topology solutions.
It may be the time to pass over from gossip about spacetime foam
and quantizing gravity to practice.

The theory proposed is natural  from viewpoint of interactions unity.
Gauge invariance and duality are ideas underlying. Some of these ideas
are not yet exhausted in gravity and of interest to apply in the Yang-Mills
theory. But it is another topic.


\end{document}